\documentclass[conference]{IEEEtran}
\IEEEoverridecommandlockouts
\usepackage{cite}
\usepackage{amsmath,amssymb,amsfonts}
\usepackage{algorithmic}
\usepackage{graphicx}
\usepackage{textcomp}
\usepackage{comment}
\usepackage{amsmath}
\usepackage{centernot}
\usepackage{subcaption} 
\usepackage[margin=1cm]{caption}
\usepackage{lipsum}
\def\BibTeX{{\rm B\kern-.05em{\sc i\kern-.025em b}\kern-.08em
    T\kern-.1667em\lower.7ex\hbox{E}\kern-.125emX}}
\begin{document}
\title{Data-driven Job Search Engine Using Skills and Company Attribute Filters}
\author{
\IEEEauthorblockN{{Rohit Muthyala}
\IEEEauthorblockA{New York University\\
New York, NY\\
Email: rrm404@nyu.edu}\\   
\and
\IEEEauthorblockN{Sam Wood}
\IEEEauthorblockA{Stanford University\\
Stanford, CA\\
Email: swood95@stanford.edu}\\
\and
\IEEEauthorblockA{Yi Jin, Yixing Qin, Hua Gao, Amit Rai\\}
\IEEEauthorblockA{EverString Technology\\
San Mateo, CA\\
Email: data-science@everstring.com \\ [0.1cm]}}
}
\maketitle
%
%
\begin{abstract}
According to a report online \cite{b34}, more than 200 million unique users search for jobs online every month. This incredibly large and fast growing demand has enticed software giants such as Google and Facebook to enter this space, which was previously dominated by companies such as LinkedIn, Indeed, Dice and CareerBuilder. 
Recently, Google released their ``AI-powered Jobs Search Engine", ``Google For Jobs'' \cite{b30} while Facebook released ``Facebook Jobs" within their platform \cite{b32}.  These current job search engines and platforms allow users to search for jobs based on general narrow filters such as job title, date posted, experience level, company and salary. However, they have severely limited filters relating to skill sets such as C++, Python, and Java and company related attributes such as employee size, revenue, technographics and micro-industries. These specialized filters can help applicants and companies connect at a very personalized, relevant and deeper level. 
\par In this paper we present a framework that provides an end-to-end ``Data-driven Jobs Search Engine". This consists of comprehensive search filters including user skill set-focused attributes and various company attributes. In addition, users can also receive potential contacts of recruiters and senior positions for connection and networking opportunities.
\par The high level implementation of the framework is described as follows:\newline 1) Collect job postings data in the United States, \newline 2) Extract meaningful tokens from the postings data using ETL (Extract, Transform and Load) pipelines, \newline 3) Normalize the data set to link company names to their specific company websites, \newline 4) Extract and ranking the skill sets, \newline 5) Link the company names and websites to their respective company level attributes with the EVERSTRING Company API, \newline 6) Run user-specific search queries on the database to identify relevant job postings and \newline 7) Rank the job search results. 
\par This framework offers a highly customizable and highly targeted search experience for end users. This framework also enables deeper analytics on the job market as a whole by providing many advanced segmenting dimensions vis-a-vis the skills-based semantic data and company level attributes.
\end{abstract}


\section{Introduction} 
Does a company utilize certain types of skills more than others? Does an individual need to have one set of skills for Fortune 500 companies and another set of skills for small-sized firms? What type of skills are required for an individual to get into a particular company or an industry? How does an individual possessing certain skills and preferences for specific industries and technologies, search for jobs? Understanding the skill sets required for a job is of importance to professionals as it governs their employability as well as their decisions to seek professional certifications. This study evaluates the skills required for all types of jobs by analyzing a broad set of online job descriptions, gathered from company websites and from Indeed's API.  From the descriptions we clean, extract, and rank the various skills required for a particular job using unsupervised algorithmic approaches.
By combining job posting data with EverString's proprietary database of company attributes, the search interface offers much more flexibility to cater to the user's personal preferences. For example, we are able to support the following real world use case: ``Find job opportunities at companies with fewer than 200 employees, whose revenue is more than 2 Million USD, within the 'digital marketing' micro industry, requiring experience in Scala or Python and knowledge of Marketo, JQuery, and CRM software". As a finishing touch, we are also able to provide the contact information for the relevant recruiters and managers at the respective companies.
\par
There are many companies like Indeed, Monster, LinkedIn, WayUp, DirectEmployers, CareerBuilder, SimplyHired, Glassdoor and Facebook which are active players on the job search market. Indeed claims to have more than 200 million unique visitors every month and Google has recently entered the recruiting space with its new AI powered job search engine - ``Google for Jobs", which displays listings from most of the job search sites mentioned above. These search interfaces, however, lack the rich skill set and company attribute preferences that job seekers need to find the best jobs matched to their personal preferences, their capabilities, and their interests and aspirations.  
\par
The framework in this paper leverages text mining and Information Retrieval techniques to transform unstructured text into useful structured data from which interesting relationships can be extracted. From job postings, we extract and rank skills by relevance, starting with TF-IDF weighting and then adjusting the results using topical information based on the job title of the posting. Encouraging results are presented after applying these techniques to a corpus of job postings extracted from Indeed and company websites.
\par
In addition to providing more nuanced search capabilities, the framework links features and attributes between the semantic data contained within the job postings to those of the company entities themselves, allowing aggregations and analytics to answer interesting questions on job markets as a whole. For instance, we can compare the top skill set requirements between small and large-sized firms (e.g. based on revenue, number of employees, and total funding raised) within a particular industry or subsector, or we can identify which subsectors are most aggressively hiring for certain positions or skill sets.
\par
This work introduces a methodology to extract and rank skills from job postings, and demonstrates the usefulness of linking advanced company attributes related to each job posting. Leveraging Everstrings's API to provide additional contextual features at the company level enhances the user experience and search capability to identify targeted and best fit job opportunities across a wider array of dimensions than is traditionally offered.
\section{Related Work}
Two widely popular classes of keyword extraction techniques were considered for extraction/ranking of skill sets in the job postings.  One class of keyword extraction/ranking technique is based on keyword matching or Vector Space models with basic TF-IDF weighting [1]. The TF-IDF weighting is obtained by using only the content of the document itself. Then several similarity measurements were used to compare the similarity of the two documents based on their feature vectors [2].  The other class of keyword extraction/ranking technique is based on using context information to improve keyword extraction. Recently, there has been lot of work on developing different machine learning methods to make use of the context in the document [3],\cite{b28}. Zhang et al. [4] discusses the use of support vector machines for keyword extraction from documents using both the local and global context. There are number of techniques developed to use local and global context in keyword extraction [3], [4], [5].

We also considered techniques used to enhance information retrieval using concepts of semantic analysis such as ontology based
similarity measures [9], [10]. In these approaches the ontology information is used to find similarity between words and find words even if the exact match is not available.  Other ways in which semantic information is extracted is
using Wordnet libraries. Wordnet based approaches have used concepts such as relatedness of words for information
retrieval [11]-[14].\cite{b27} Demonstrated that informative structured snippets of the job postings can be generated in an unsupervised way to improve the user experience for a job search engine.  Croft et al. in his book [15], describes the various uses of search engines in information retrieval.  Recent works [16] have shown the use of encyclopedic knowledge for
information retrieval. Lian et al [17], describe the use of Google distance to find concept similarity. Google distance
based approaches have been used in various applications such as relevant information extraction [21], [18], keyword
prediction [19], and tag filtering [20]. Given the keywords that we need to look for and rank them based on the documents given is also a similar problem.

After an extensive search of the literature, we found limited references on skill set-based approaches to help job seekers to search for the best fit jobs.  Most of the job hiring websites today rank the job postings based on the keywords given by the user. The ranking is done considering the entire corpus of documents and the key words given, which limits the users to a single level search. \cite{b24} has done very interesting work using the skill set where a user can search for the jobs based on the skills but this search is still single level where the job postings are ranked using the keywords given by the user.
  
Literature on fostering college graduates to make them more employable is well grounded and has a rich history  in many fields of study. Birch, Allen, McDonald, and Tomaszczyk (2010) studied the considerable  mismatch  between student job expectations and experiences  with  what the business community expects.\cite{b25} discusses the need for accurate skills assessments of employees in large, global, client-facing enterprises and shortcomings of existing systems for obtaining and managing expertise. Gault, Leach,and Duey (2010) examined the differences between how much training students  believe  they  need  versus  what  employers  already  expect  from  the  students  when  hired. These sources definitely note a mismatch in expectations and skills. Tanyel,  Mitchell,and  McAlum (1999),  in  surveying  prospective  employers  and  university  faculty  on  their perceptions of the skills and abilities business school graduates should possess in to compete in industry, found that about half of the skills listed by each group differed. In other words, what faculty thought were the  important  skills  versus  what  prospective  employers actually  sought  did  not  match.  To identify the mismatch between the skills expected by employers and those possessed by job seekers,\cite{b26} profile job titles by effectively quantifying the relevance of skills.

Here we propose a method to get this search to a multi level where a user can search for jobs based on different categories like technologies, skills, revenue, company size, industry and many more and the search results are ranked taking a lot of features into consideration, such as Alexa ranking (measures the web traffic for websites, i.e. popularity of the company), social followers, revenue, user given keywords (skills, industry, technologies) and many more. 

\section{Design and Implementation}
\begin{figure}[h!]
\begin{minipage}[b]{1.00\linewidth}
\centering
\includegraphics[width=9cm,height=6.5cm]{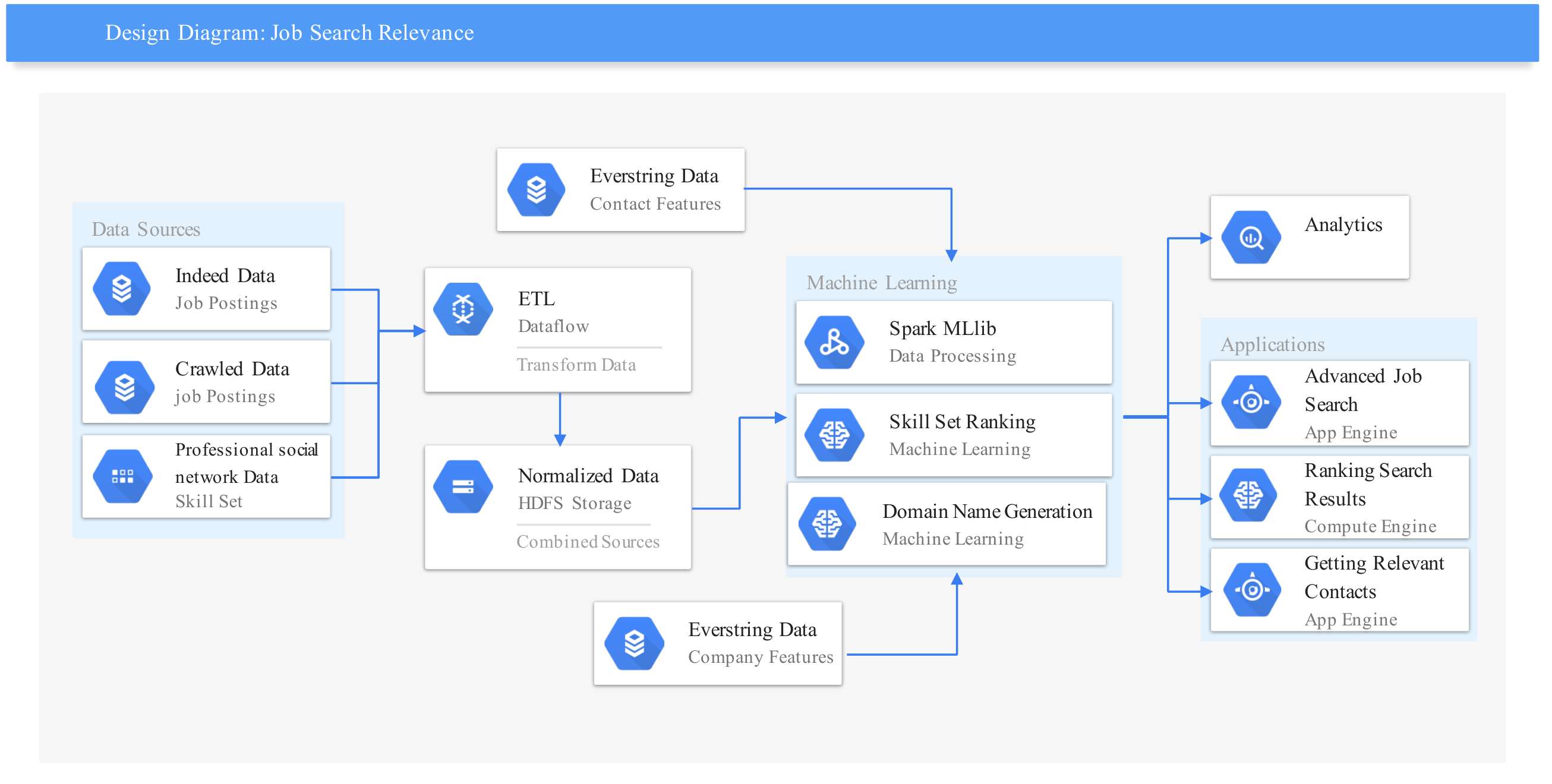}
\end{minipage}
\centering
\caption{\small Design Diagram - Job Search Engine}
\label{fig:design-diagram}
\end{figure}

At a high level, our system deals with text associated with job postings online, to infer and rank the skill set for each job posting using the job descriptions  and the job title. The skill set extracted for each job posting is joined with the Everstring's knowledge base to enhance it with the company features. The combined data is used to search for the jobs with a very specific search queries given by the user/job seeker. The search results are finally ranked using relevance score, will describe in \ref{QRR}. Some interesting analytics which are useful in the job hiring market are developed and shown in this paper.
The database generated can be used to find the job postings which exactly match the users' requirement and also give the users with the contacts of the relevant people currently working in the companies of the corresponding job postings.
\newline
The process is described below.
\begin{enumerate}
  \item Extract the job postings and skills data from indeed, company job pages and professional social networks\cite{b23}.
  \item Extract useful tokens of data and store them in distributed database - ETL(Extract Transform and Load).
  \item Normalize, clean the data as described in \ref{BB}. Obtain the domain name from company name as described in \ref{CD}. 
  \item Extract and rank the skill set required for a job as described in \ref{TG} and \ref{WG}.
  \item Merge with the Everstring Data using API to extract useful Insights.
  \item Run user specific search queries on the database to receive relevant job postings in a ranked order \ref{JSR}.
  \item Perform analytics \ref{Analy}
  
\end{enumerate}

\subsection{Data Extraction}\label{AA}
We constructed a web crawler to crawl the job pages of companies to extract the company names, urls, job descriptions, job titles and company address. This was done by exploiting regularities in the html structure of the companies website and by constructing extracting functions to automatically extract company names and descriptions of jobs from arbitrary websites. Additionally, we used Indeed's API to fetch the job postings with all the necessary features.
Here, we didn't crawl the entire job posting web page, instead we crawl only the useful information that is required; such as job description, titles, company name and address. This is in contrary with the traditional web crawlers which crawl the entire website and later extract the useful features from the entire unstructured website data. This crawl was done by using an XML based approach and viewing the task of data extraction as a multi-step process where the goal extends far beyond simple``screen scraping" as mentioned in \cite{b7}. Navigation rules and extraction rules were optimized by hand to produce semi structured data.  
The skill sets mentioned in the professional social networks are extracted by  leveraging the Linking Open Data project DBpedia\cite{b23} to generate a set of 750k unique skills.

\subsection{Data Preprocessing}\label{BB}
\subsubsection{Skills Data}  \label{PSN}
By generating the co-occurrence counts of the skills \cite{b23}, we can filter the skills based on a count threshold which is decided based on the frequency and the number of words in the skill. We then generate lower cased skills and apply basic normalization techniques on the skills, using several common patterns. For example replacing ``\&" or ``+" with ``and" unless ``r\&d", removing the dash in ``e-mail" and running basic lemmatization on the skills. While normalizing, we keep the original lower cased skills and the normalized skills in a sequence. For example, if the input is ``e-mail", we generate a sequence of the original skill and the normalized skill - Seq(``e-mail",``email"). We then generate a lemmatized skill (standard word in WordNet) using the lower cased skill. To generate the lemmatized skill, we parse the skill into its part of speech(POS) to obtain the word Ngrams of the skill. We then replace any Ngram with the one that is present in WordNet.  If the Ngram is not present in WordNet the Ngram is not replaced. Using the lemmatized Ngrams, we combine the Ngrams back to generate the lemmatized skill. Leveraging Everstring's Human Intelligence Tasks (HITs) system, an in-house on-demand workforce platform, we filter the 1000 most frequently occurred lemmatized skills. For example, words like ``team", ``knowledge",``information"  which occurred more frequently in the dataset and are removed from the skill set as they act as outliers in our skill set.

We construct a lemma dictionary using the lemmatized skills and its corresponding sequence of normalized and lower cased skills. The lemma dictionary is a map from the normalized and original skill to the lemmatized skill. The normalized and original skill will be the key in the map. The sequence key is split to generate two different keys. For example, Seq(``A",``B") map to ``C" becomes a map from $``A"\rightarrow``C"$and $``B"\rightarrow``C"$. This generates a dictionary that maps the original and normalized skills to the lemmatized skills. 
The total number of distinct lemmatized skills is reduced from 750k to 73k after filtering and lemmatization of skills.
\newline
For Example: ``c\#/.net", ``c\# / .net",``c\# \& .net" and ``c\# and .net" are mapped to ``c\# and .net". ``systems installations",``system installations",``systems installation", and ``system installing" are mapped to ``system installation"
\subsubsection{Job Postings Data}
In this section, we discuss multiple information extraction (IE) steps performed on the job postings data. All IE steps are performed on the raw text of the job titles, company names and job descriptions to retrieve useful and structured information and to generate and rank the skill sets to take the job search process to a much finer level.
\begin{itemize}
\item \textit {Title Normalization and Parsing:} \label{TN}
The title string in its natural language form is first normalized and then parsed to obtain the management level and the department(s) it falls under. The title normalization steps are: \newline
\textbf{1.} Replace specific characters such as spacing out commas and handling special characters.\newline
\textbf{2.} Remove level information, e.g. strip out roman numeral tails, ``level 1", etc. Apply Ngram substitutions, e.g. $VP \rightarrow Vice \space President$; $e- mail \rightarrow email$.\newline
\textbf{3.} Add acronym forms with periods after each character, e.g. $ceo \rightarrow c.e.o.$ 
\par At a high level, we apply character level(1), sequence level(2) and token level(3) transformations to normalize the title. In between these steps we use regex to split the title at different levels to identify the title Ngram/tokens. 
We parse the normalized title to obtain its management level and department(s).  The title is classified into one of five different management levels (C-Level, VP-Level, Director, Manager, Non-Manager) and one or more departments (``Administrative", ``Computing \& IT", ``Engineering", ``Educator", ``Finance", ``HR", ``Marketing" and many others).  This parsing is performed using a manually generating mapping from title Ngrams to management levels and departments.
\end{itemize}

\begin{itemize}
\item \textit {Company Name Normalizer:} \label{CN}
The company name string in its natural language form is normalized to a form that facilitates matching to a website.\newline
\textbf{1.} Convert the company name to lowercase and replace ``\rq s" with ``s", e.g., $``Macy's" \rightarrow, ``macys"$, and convert non-alphanumeric characters to spaces.\newline
\textbf{2.} Generate a set of suffixes and prefixes that are commonly used in company names like, ``llc", ``ltd", ``corp",  ``inc", ``co" and many others and use this set to drop the occurrences of these words in the company name. \newline 
\textbf{3.} Generate a set of stop words that are not useful to recognize a company such as, ``technologies", ``management", ``service", ``pvt", ``group", ``solutions" and many others. Use this set to drop the occurrences of these words in the company names, irrespective of their position. This method is inspired by the CompanyDepot\cite{b8} system.
\end{itemize}

\begin{itemize}
\item \textit{Job Description Cleaning:}
To normalize the job descriptions, we use the same normalization schema as the one used to normalize the skills data.
\end{itemize}

\begin{itemize}
\item \textit {BOW Generation Using Skill Sets:}
The lemma dictionary which maps the lemma to the original and normalized skill is constructed using the skills mentioned in professional social networks. We use this lemma dictionary to count the occurrences of the normalized or the original skill in the job descriptions and generate a map from the lemma skill to its count, keeping the track of the original form of the skill. Here, we create a map from the lemmatized skill to the struct of the total lemma count and the map of original or normalized skill Ngram and its count. 
\end{itemize}

\subsection{Company Name to Website}\label{CD}
In order to utilize company firmographic APIs that are available on the web, we had to convert the normalized company name associated with each job posting to their respective company website.  To perform this conversion, we leveraged EverString's Company Name to Website service which takes in inputs parameter such as name, location (parsed or unparsed street, city, state, zip) and outputs the website associated with the input with a confidence score.  EverString's Company Name to Website Service was generated by triangulating both crawled results as well as multiple purchased data sources to generate a comprehensive alias table for fast query.  Using this service, we were able to convert company names like ``Amazon Drive", ``Amazon Web Services", ``Amazon Prime", ``AmazonFresh", ``Amazon HVH",``Amazon Corporate LLC", ``Amazon Logistics",``Amazon Web Services, Inc", ``Amazon.com.dedc, LLC",``Amazon Fulfillment",``Amazon Fulfillment services", ``Amazon" to the company's website amazon.com with high confidence scores.

\subsection{TF-IDF Generation}\label{TG} 
\textbf{Document Frequency Generation:} 
We begin by counting document (here a document refers to a job posting) frequencies for all unique skills (lemmas, original and normalized skills) present in the skills lemma dictionary computed in \ref{PSN}. We compute document frequencies for skill lemmas and filter out lemmas that do not satisfy a minimum document frequency threshold. For the remaining valid lemmas, we add the original Ngram forms (if different) to the list of document frequency counts, so we have a unified dictionary which maps all the skills in the documents to its count. 
Documents with zero document length (sum of all the counts of the skills - \textit{docLen}) are removed from the corpus. Reserved keyword ``nDocs" is set to represent the number of valid documents in our corpus which helps in computing the inverse-document-frequency later.
\newline
\textbf{LTU term weighting scheme(TF-IDF):}
The term frequencies are generated for each document/ job posting by counting the number of occurrences of the lemma skill as described in the ``BOW generation" section above. The weighting for each skill in a document is generated using the following formula:\newline
\[ LTU =\frac{(\log_2 \left(tf \right) + 1) \log_2 \left(\frac{nDocs}{df} \right)}{(0.8 + 0.2 \frac{docLen}{avgDocLen})} \]
where, \textit{tf}(Term Frequency) is the number of occurrences of this term (the skill for which we are computing the weight) in the current document, \textit{docLen}(Document Length) is total count (with repetitions) of terms present in the current document, \textit{nDocs} (Number Documents) is the total number of documents (job postings) in the database, \textit{avgDocLen} (Average Document Length) is the average number (with repetitions) of terms present across all documents, \textit{df} (Document Frequency) is the number of documents containing this term. LTU weighting uses a pivoted document length normalization scheme, based on the intuition that words that appear the same number of times in shorter documents should be considered more relevant than in longer documents. This weighting schema also incorporates sub-linear term frequency scaling.
Finally, the scaled TF-IDF is generated for each document by scaling the maximum TF-IDF value of a skill to 1.

\subsection{Weights Generation}\label{WG}
Ranking the skills using just the TF-IDF score did not produce good results as we have only taken the job descriptions into consideration for generating this score. While the job description contains an extensive amount of information about the company, the benefits they provide is mixed.  The primary concern is much of the information acts as a noise in our system. This drives us to include one more additional dimensions for a better ranking. In order to achieve this, we use the job title to generate a weighting schema on top of the TF-IDF that we have computed previously. Intuitively, we want a skill to be weighted higher if it occurs more frequently under a title Ngram. If a particular skill is very common in the dataset then it should be weighted lower. For example, job postings with ``Software" in the title will weigh the skills like ``C",``C++",``Java'',``OOP'' higher as they frequently occur in the job posting with ``Software'' in their title.  They will not occur in many non technical job postings (a high percentage of job postings in our corpus), which boosts the probability of having these skills given ``Software'' in the title.  In this way, we have added one more dimension to weigh the skills by considering the titles of the jobs. The weighting schema is mathematically explained using a formula which is explained in the following sections:


\subsubsection{Title Ngram Generation}
Initially, we normalize the title and remove all  punctuations present in the title. We then split the title to generate a set of all unigrams present in the titles. This set is filtered based on the occurrence frequency and stop words. We then generate bigrams from the title by looking at the sequence of two words in the original normalized title and filtering out the bigrams by checking if it is present in the original normalized title. Eg: If we have ``Big Data Software Engineer" as a title, we look for all bigram sequences (``Big Data",``Data Software" and ``Software Engineer") in the title and add the bigrams that are present in the title to generate the final list of title Ngrams. In this example, ``Software Engineer" will be added to the title Ngrams list. After generating this list, we split the normalized title to obtain the unigrams, bigrams and filter out the grams that are not present in the title to obtain the title Ngram list for each job title.

\subsubsection{Weight Matrix Generation}
Now that we have an array of title Ngrams for every title in the corpus, we use this information to rank the skills that are extracted for each job posting (`` BOW Generation'' in the section above) by generating a weight for every skill present in a job posting, using a count matrix between the skills and the title Ngrams across all the documents. 
The count matrix\cite{b22} is computed by generating the term frequency of a particular skill present in all job descriptions that has a particular title Ngram in its title. E.g. we obtain a count array (of length equal to number of unique skills) for every unique title Ngram. This generates a matrix giving the count of the occurrences of all skills corresponding to a particular title Ngram. Using this count matrix, we compute the weight for every skill in a job posting by adopting the following formula:
\[ weight(skill) =\frac{prob(skill \vert title \space Ngram)}{prob(skill)} \]
The above weighting formula supports our intuition that the skill which is more common across all the documents will be weighted lower and the skill is weighted higher if the probability of finding it in all the job descriptions which has a particular title Ngram is higher.
For a particular skill in a job posting we obtain a different weight for each title Ngram of a job posting because each title in a job posting has multiple title Ngrams. To generate a single weight for each skill in a job posting, we average all of the weights computed for this skill and all of the job title Ngrams. 

\subsubsection{Final Score Generation} \label{FSG}
In the previous section we computed the weights for all of the skills in a job posting. This weight is multiplied with the TF-IDF/LTU computed previously to obtain a score for a particular skill in a job posting. This score and many other external factors about the company are used to rank the skills corresponding to a job posting. 

\subsection{Query Results Ranking} \label{QRR}
After processing the query the search results are ranked based on the following formula:
\[ Rank(document) = Attribute Score * Avg(weight(skill)) \]
\textit{Avg(weight(skill))} is the average of all the weights \ref{WG} of the skills corresponding to a job posting.
As we have a weight for all the skills for each job posting, we take the skills given by the user and compute the aggregated score, which is the average of user given skills weight.
\[ Attribute Score =  feedback*af*ef*nlf*cks \]
where, \textit{af} is the alexa Factor \cite{b33}, \textit{ef} is the Employee Factor(A factor computed using the information of the number of employees in the company), \textit{nlf} is the Number of Lemmas Factor, \textit{cks} =  $\sqrt[2]{score}$ where,  \textit{score} is the micro industry keyword score and feedback is the score defined by the number of user clicks. 



\section{Results}
In this section, we will show an example of how a job seeker can search for relevant jobs based on the person's skill set, the desired industry space, the preferred technology stack and many other skill and company attributes.  The job seeker can then query for relevant contacts within the company which helps the job seeker to connect and network. We then compared our search results with other job search websites such as Indeed and Google.  

In addition, this section presents examples of enhanced analytics including analysis of the most common skill sets required by ``staffing services",``health services'' and ``restaurant chain'' industries, the most desired technologies in the US job market and presenting the top recruiting companies that are looking for prospects with specific skill sets (``Java" and ``Masters Degree"), or uses specific technologies (``Tableau'' and ``MongoDB''). 

\subsection{Job Search Results:} \label{JSR}
\textbf{Query-1:} A User with a bachelor's degree and has Python and Scala programming skills wants to search for the jobs in companies which uses jQuery technology, wants to work in the ``engineering" vertical with companies whose revenue is great than 1 Million USD (the revenue that we show in the following results are in terms of 1000 USD 
) and the number of employees to be between 50 and 200.
\par
\begin{figure}[h!]
\begin{minipage}[b]{1.00\linewidth}
\centering
\includegraphics[width=9.2cm,height=7cm]{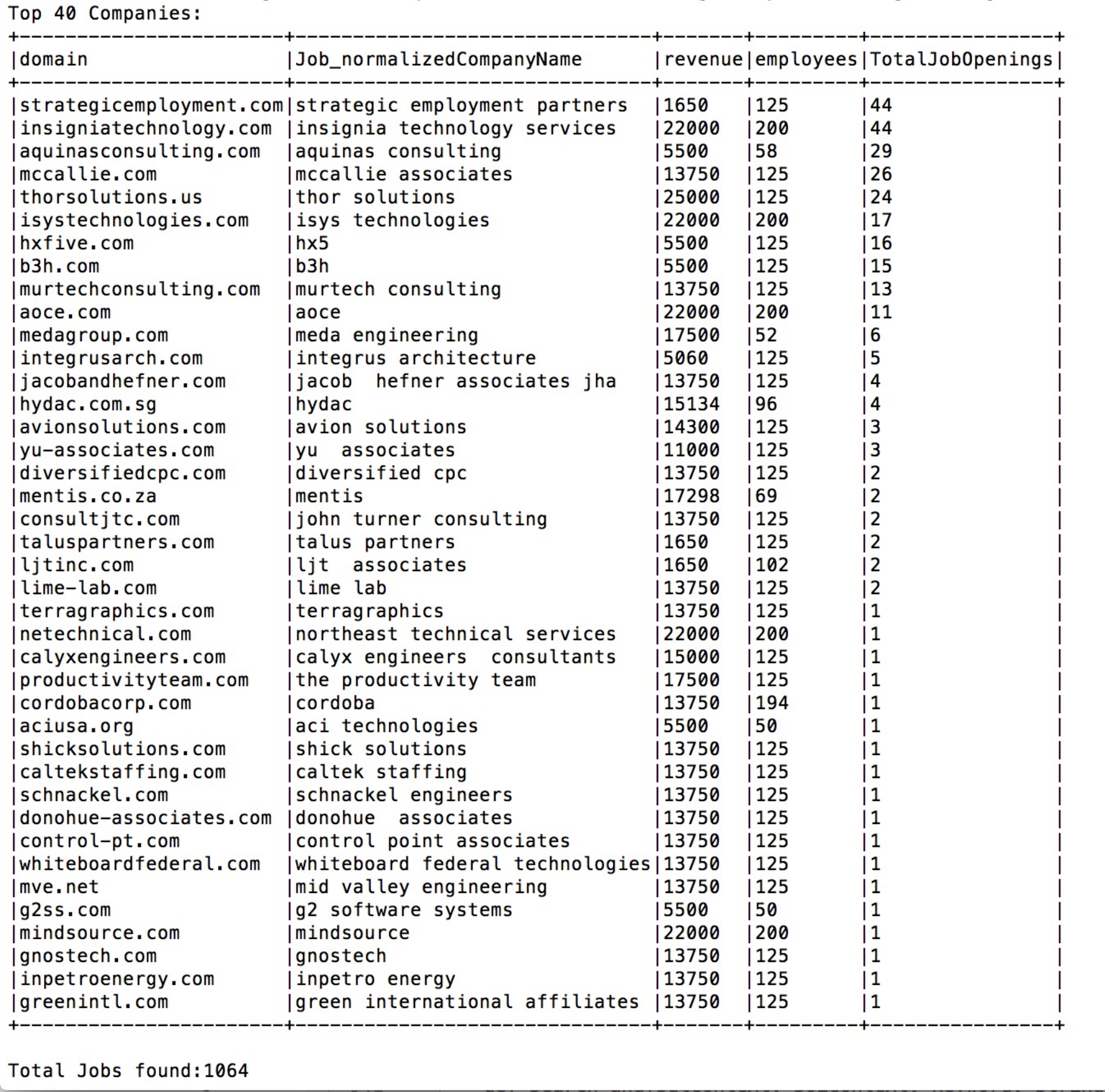}
\end{minipage}
\centering
\caption{\small Top 40 Companies Found Using Query-1 }
\label{fig:query1}
\end{figure}

The top 40 companies that we identified with the above search query are shown in Fig.~\ref{fig:query1}.

\begin{figure}[h!]
\begin{minipage}[b]{1.00\linewidth}
\centering
\includegraphics[width=9.2cm,height=7cm]{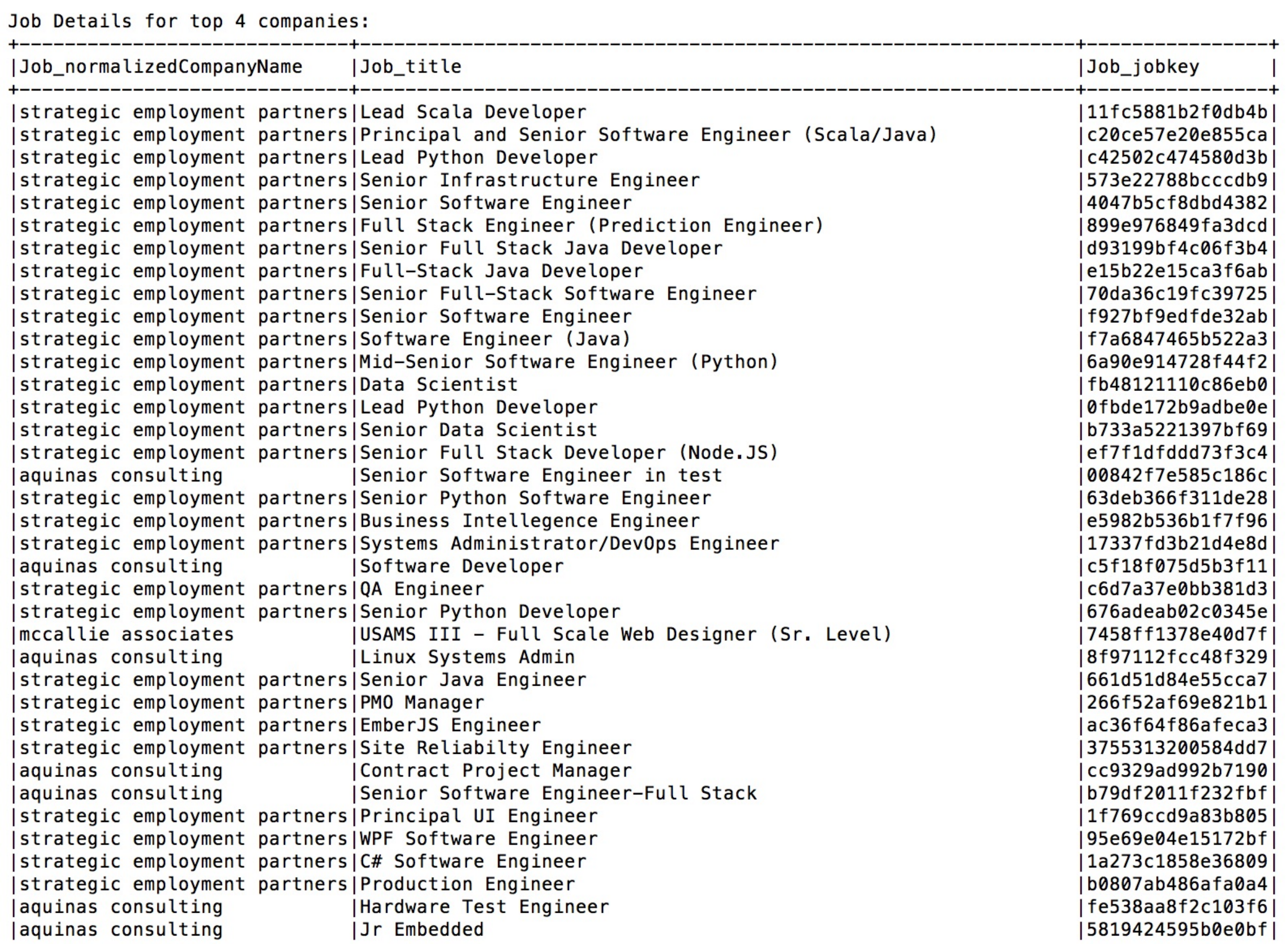}
\end{minipage}
\centering
\caption{\small Job Details for Top Three Companies }
\label{fig:query1-details}
\end{figure}

Showing detailed job results for the \textbf{top three} companies from Fig. \ref{fig:query1}, in ranked order is shown in Fig \ref{fig:query1-details}. To clarify on few results which may look odd: we have results like, ``Senior Full stack Java Developer" shown up for the above query. When looked into the job description \cite{b35} (Job Key: d93199bf4c06f3b4), we have the following statement: ``Strong professional experience with Java and at least one other language - Python, Go, Scala, etc." So this supports our search query. Here we can also show the ranked job postings based on the query, but here we want to present the results based on the company, similar to topic search engine. 

\begin{figure*}
  \includegraphics[width=\textwidth,height=9cm]{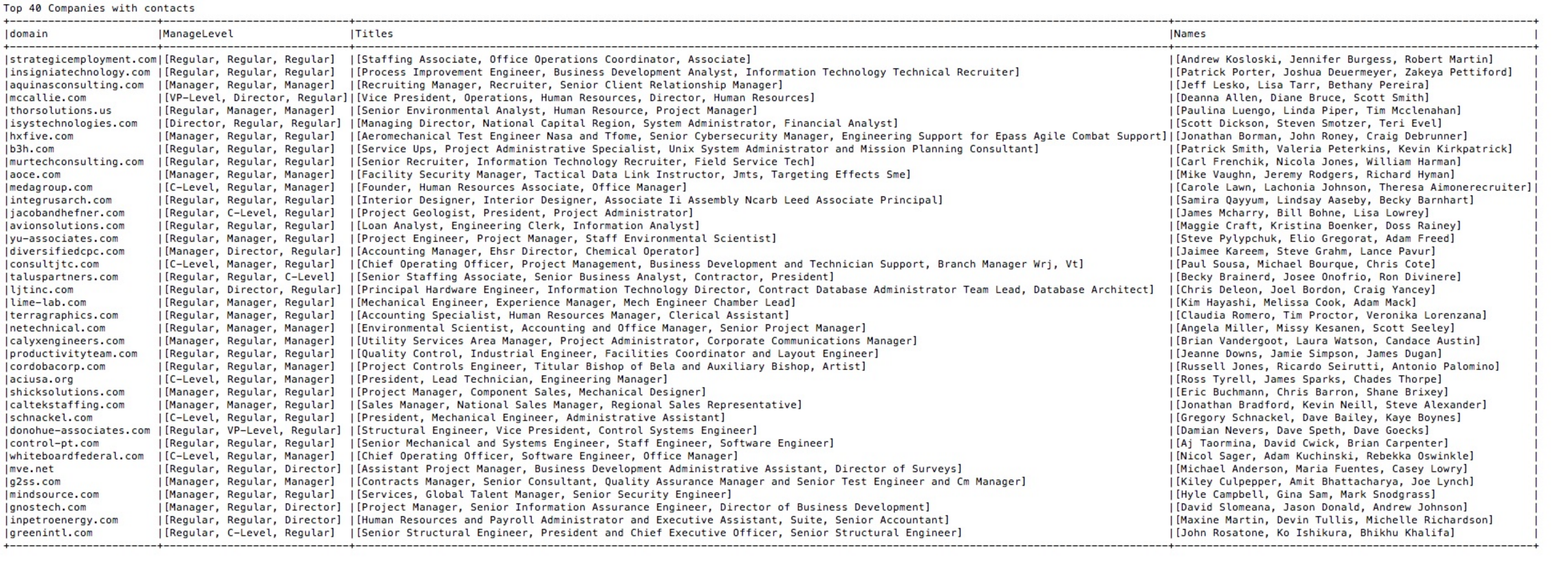}
  \caption{\small Top 40 Companies with Contacts }
  \label{fig:contacts}
\end{figure*}

The contact information of the recruiters of the top 40 companies in the Fig. ~\ref{fig:query1} are shown in Fig. ~\ref{fig:contacts}. Considering the privacy violation, we are not showing the phone numbers and the emails of the recruiters in this paper.  

No job search website supports querying with number of employees and revenue. When we searched for a combination of the five keywords (``jQuery", ``engineering", ``Python", ``Scala", ``bachelor degree") described above, we received zero results using Google's job search. Indeed's job search produced many results satisfying any one of the keywords i.e. jobs that satisfy either ``jQuery" or ``engineering" or ``Python" or ``Scala" or ``bachelor degree''.  
Our results are not exhaustive as we ran this experiment on a limited set of job postings.  However, when this approach is used on a larger dataset, we obtain much better results and with a better user experience.
\subsection{Analytics:} \label{Analy}
\begin{figure*}
  \begin{subfigure}[b]{0.23\textwidth}
  	\captionsetup{width=0.9\textwidth}
    \includegraphics[width=\textwidth, height = 10.2cm]{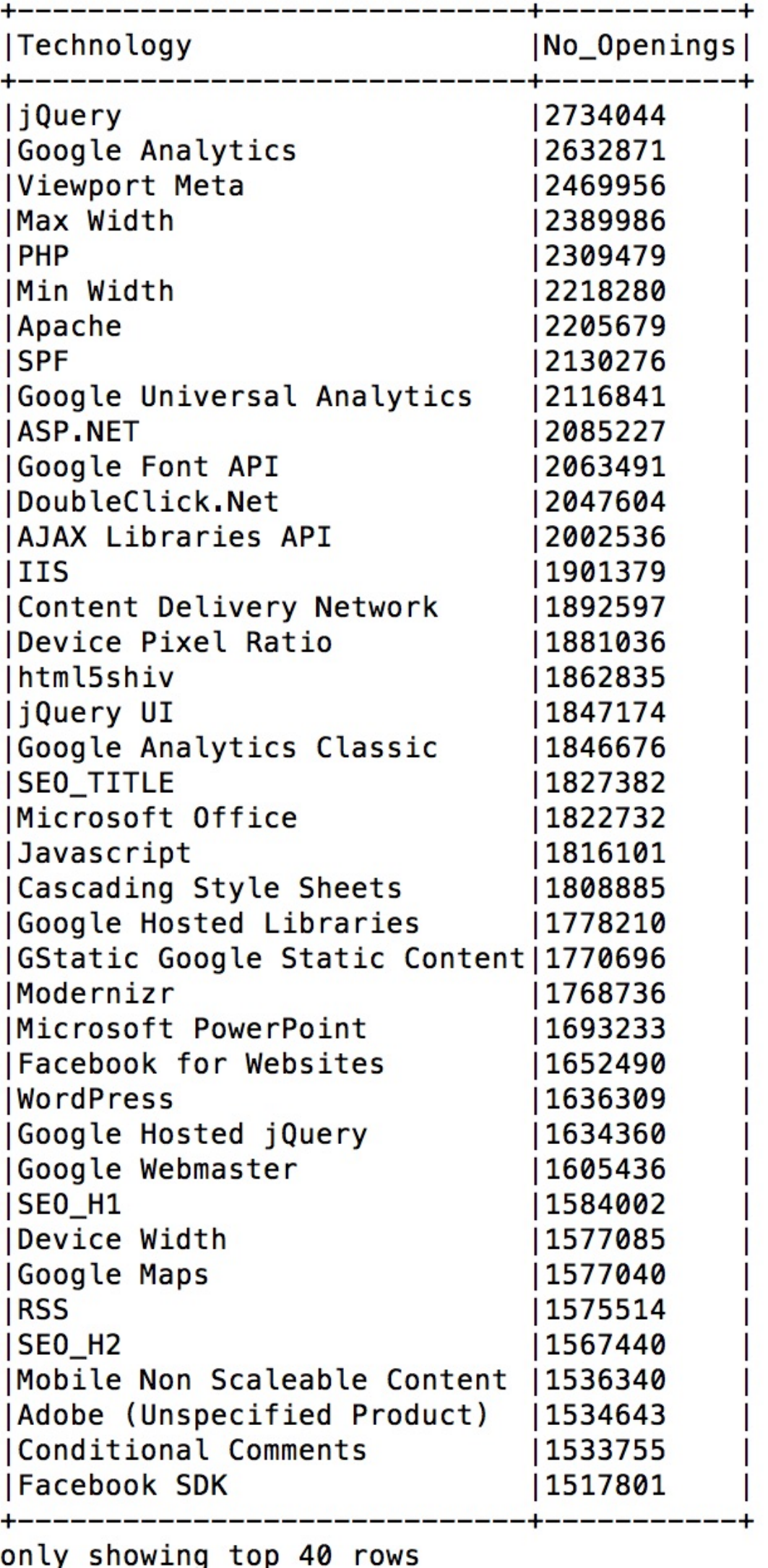}
    \caption{Top 40 Technologies}
    \label{fig:1}
  \end{subfigure}
  \begin{subfigure}[b]{0.23\textwidth}
  	\captionsetup{width=0.9\textwidth}
    \includegraphics[width=\textwidth, height = 10cm]{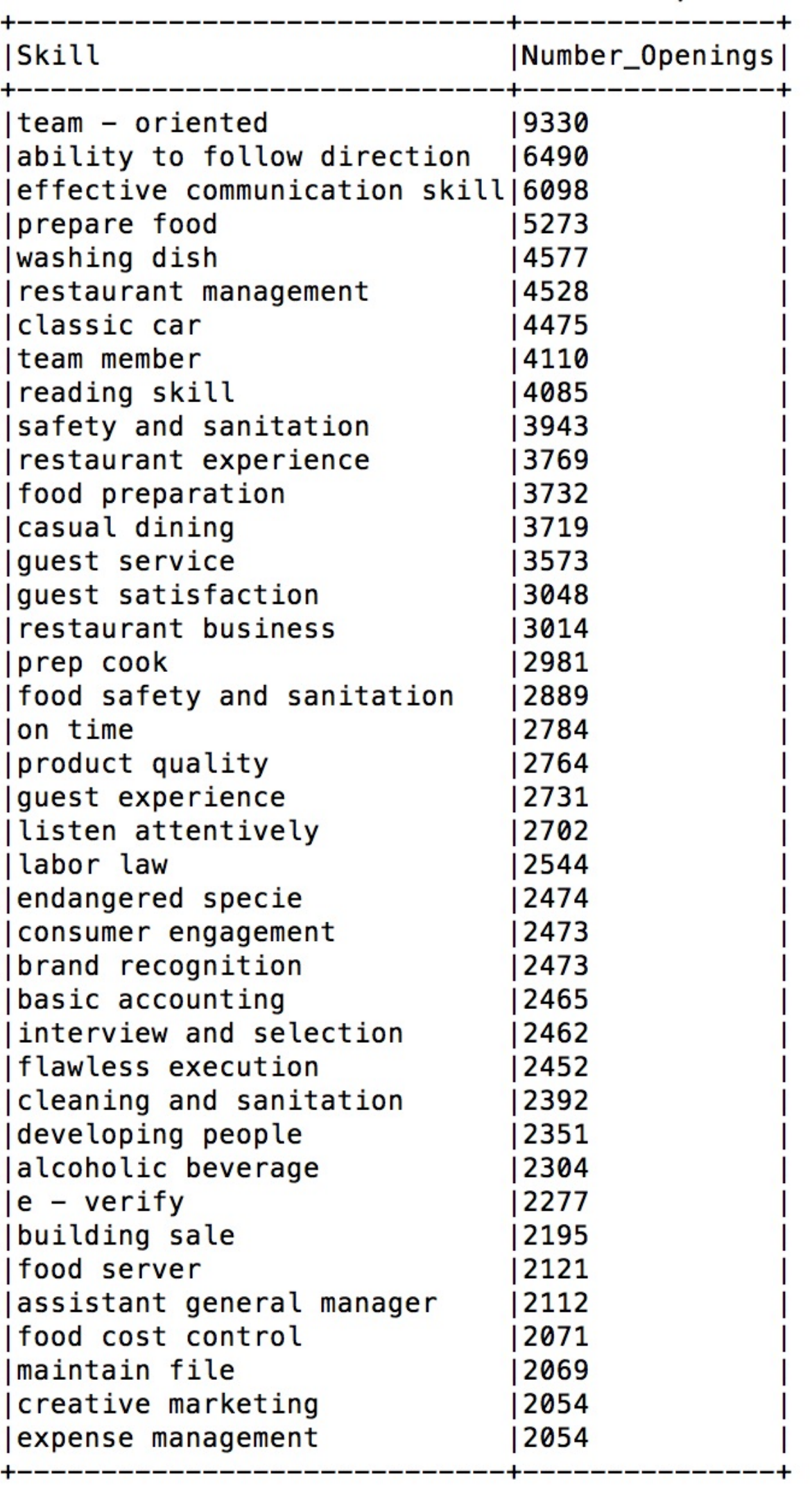}
    \caption{Top 40 Skills in the Restaurant Chain Industry}
    \label{fig:2}
  \end{subfigure}
  \begin{subfigure}[b]{0.23\textwidth}
  	\captionsetup{width=0.9\textwidth}
    \includegraphics[width=\textwidth, height = 10cm]{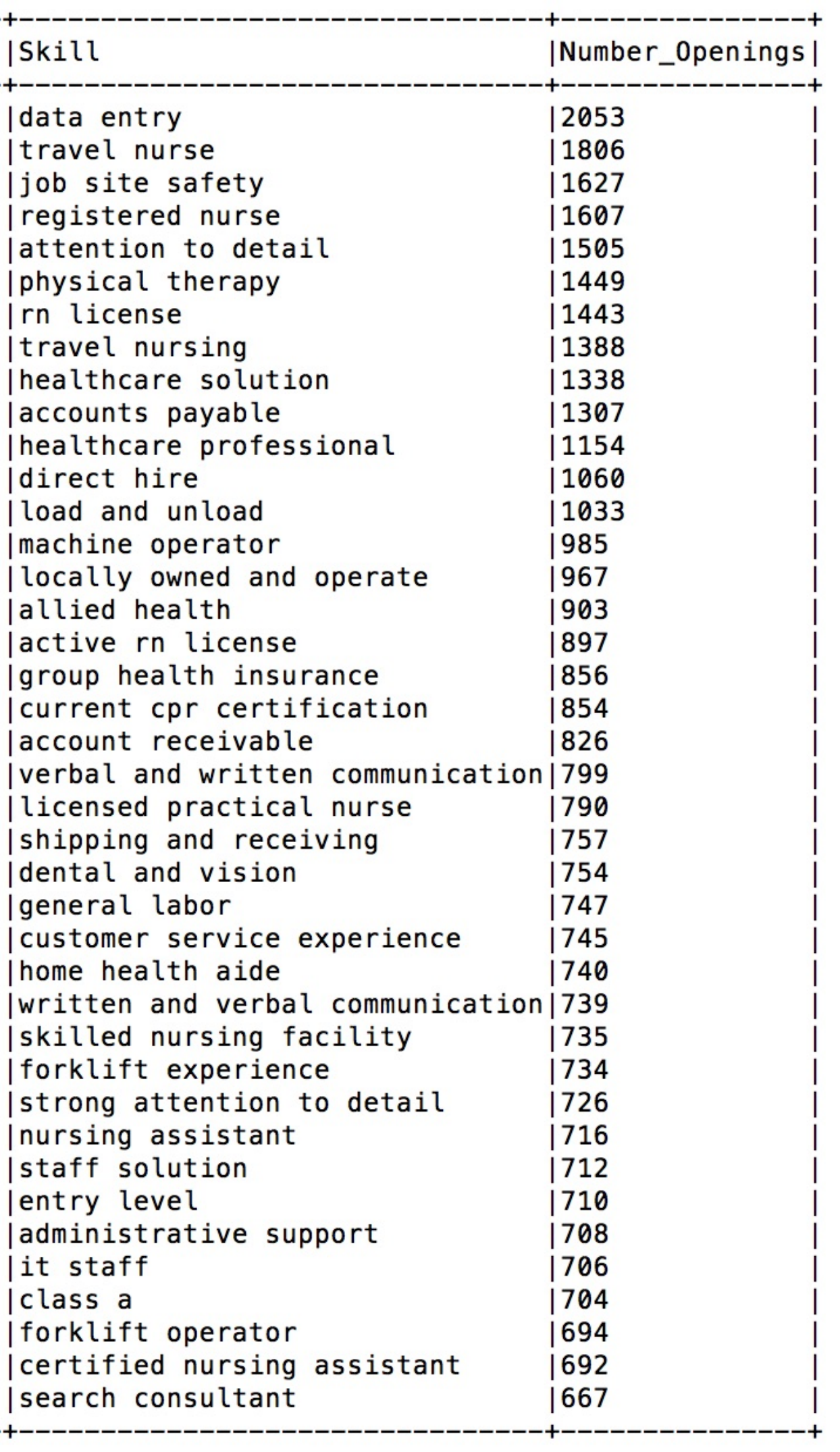}
    \caption{Top 40 Skills in the Staffing Services Industry}
    \label{fig:3}
  \end{subfigure}
  \begin{subfigure}[b]{0.23\textwidth}
  	\captionsetup{width=0.9\textwidth}
    \includegraphics[width=\textwidth, height = 10cm]{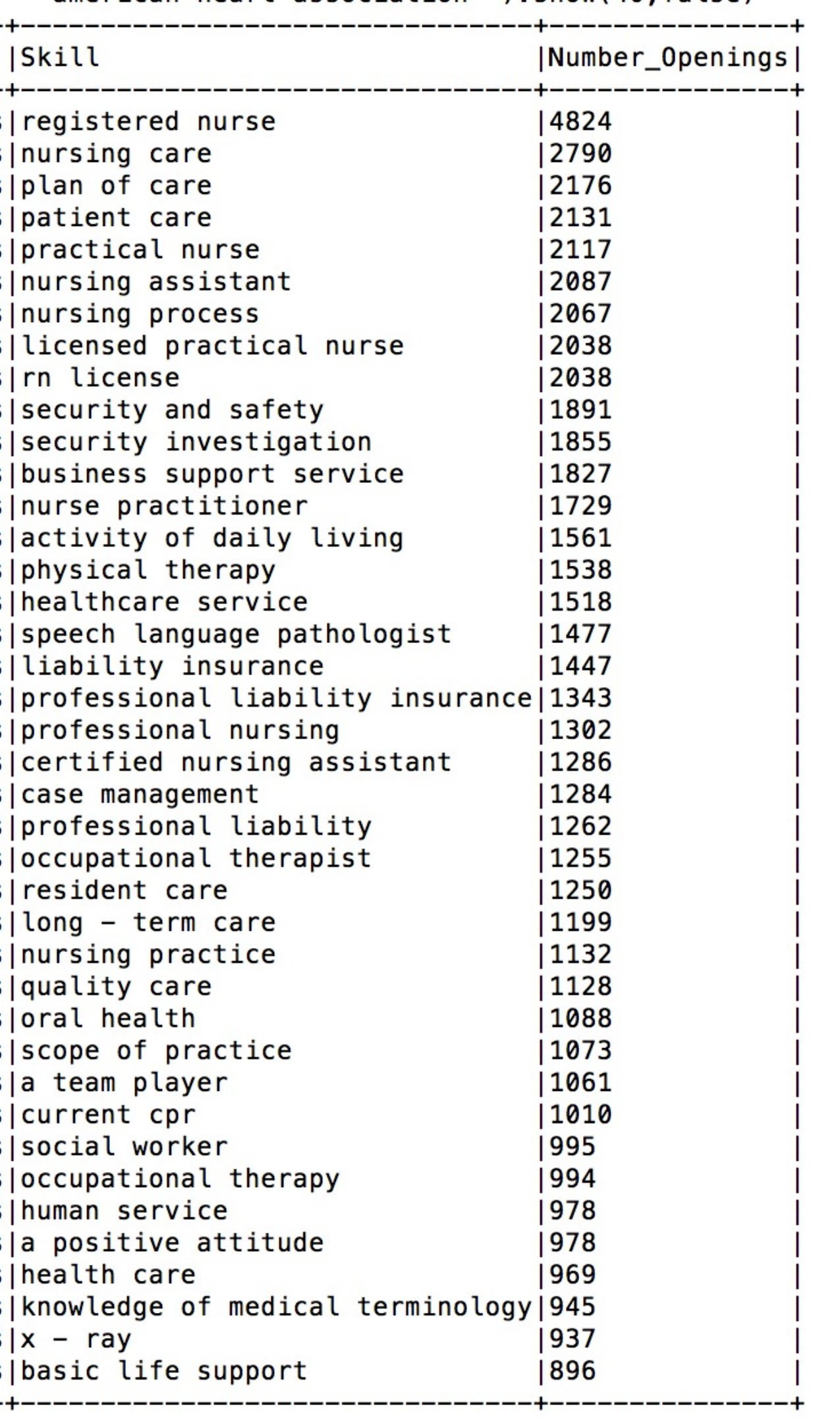}
    \caption{Top 40 Skills in the Health Services Industry}
    \label{fig:4}
  \end{subfigure}
\caption{\small Analytic -1 }
\label{fig:Analytic - 1}
\end{figure*}

\textbf{Analytic-1:}
 The top 40 technologies across all job postings are shown in Fig. \ref{fig:1}.  The top 40 skills that companies in the restaurant chain industry are looking for are shown in Fig. \ref{fig:2}.  The top 40 skills that companies in the staffing services industry are looking for are shown in Fig. \ref{fig:3}.  Top 40 skills that companies in the health services industry are looking for are shown in Fig. \ref{fig:4}

\begin{figure}
  \begin{subfigure}[b]{0.25\textwidth}
  	\captionsetup{width=0.9\textwidth}
    \includegraphics[width=\textwidth, height = 9.3cm]{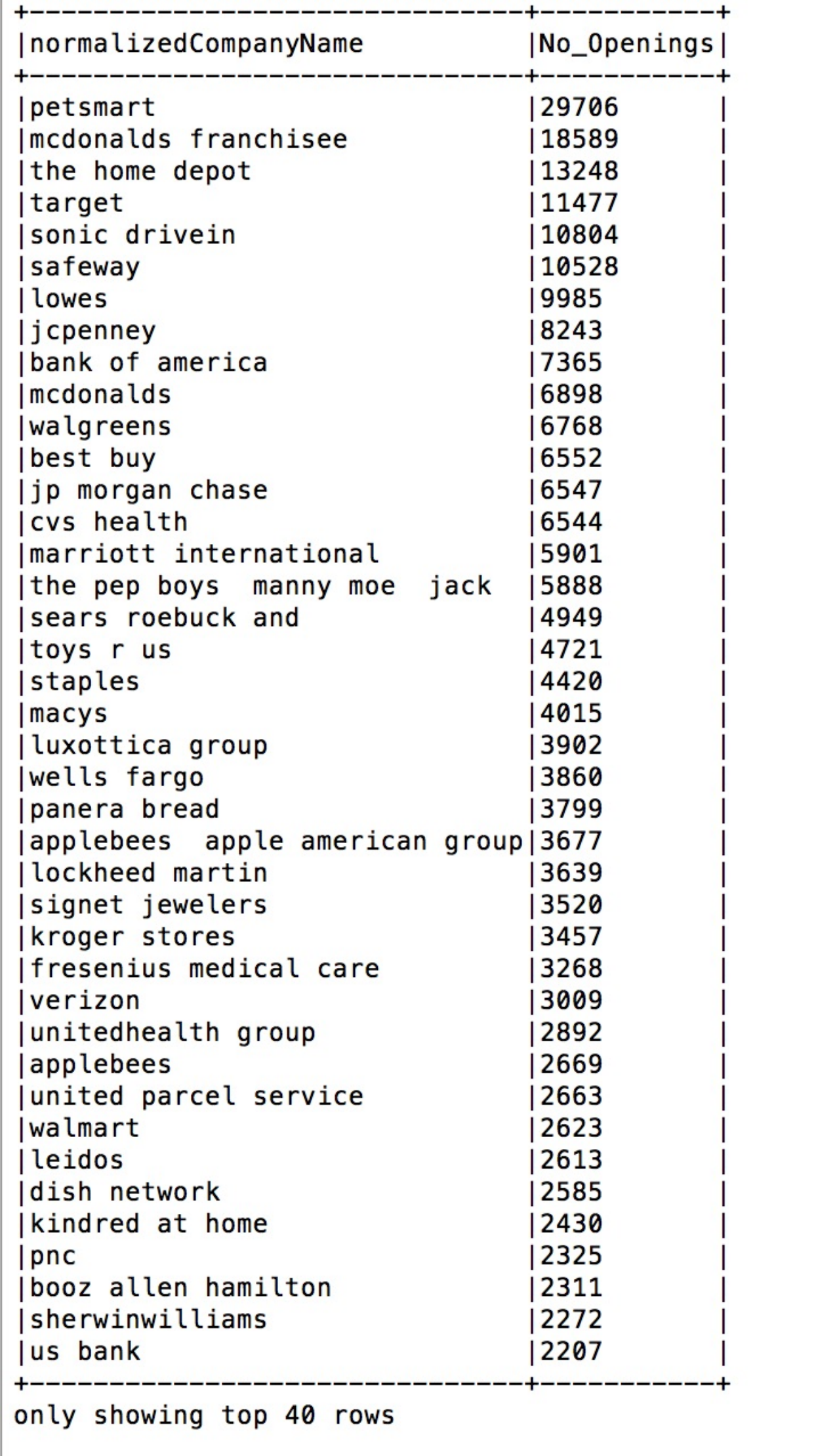}
    \caption{Top companies that uses ``Tableau'' and ``MongoDB"}
    \label{fig1:1}
  \end{subfigure}
  \begin{subfigure}[b]{0.23\textwidth}
  	\captionsetup{width=0.9\textwidth}
    \includegraphics[width=\textwidth, height = 9cm]{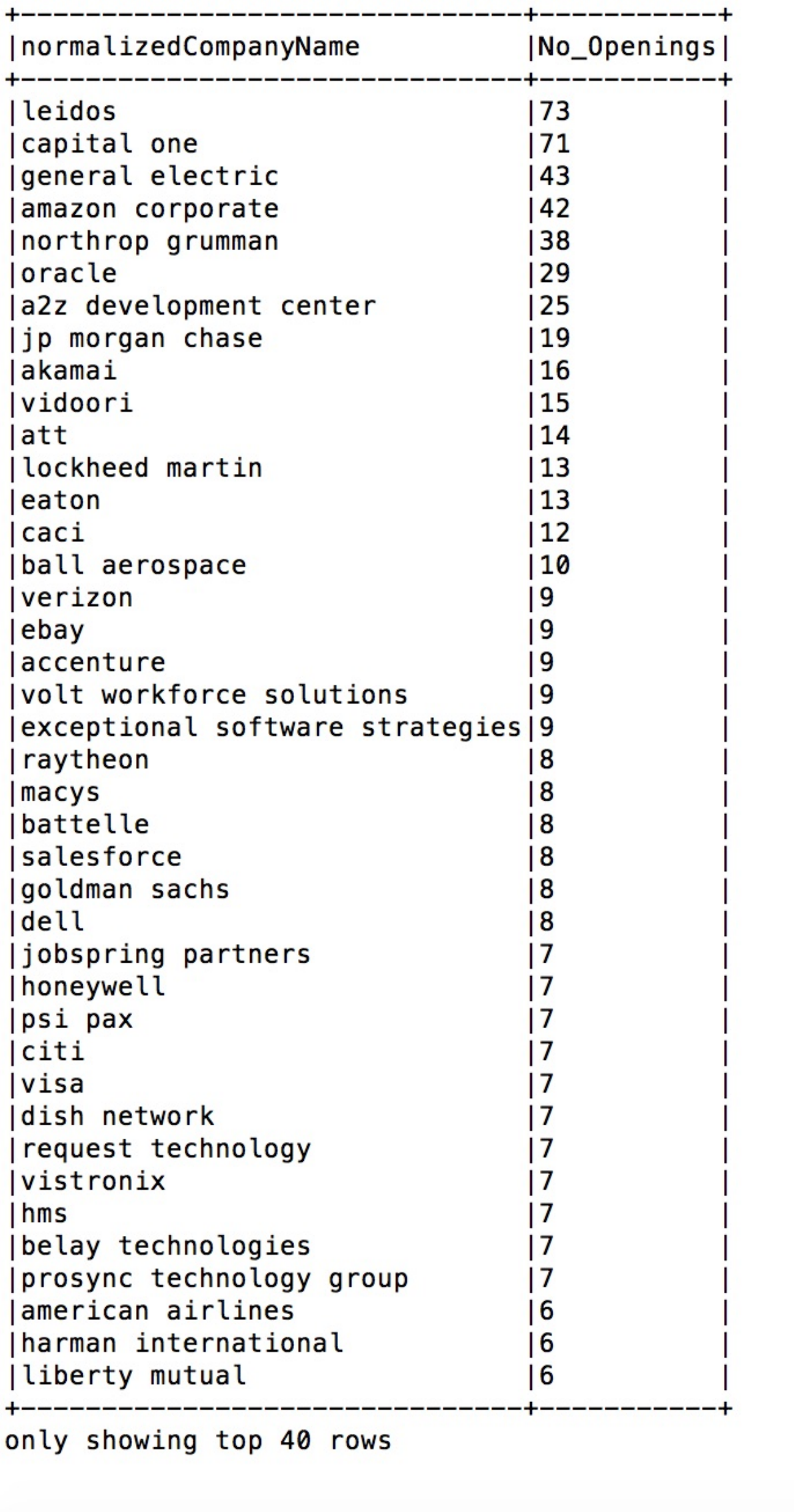}
    \caption{Top recruiters that require a person who knows ``java'' and has a ``master degree'' }
    \label{fig2:2}
  \end{subfigure}
\caption{\small Analytic - 2 }
\label{fig:Analytic-2}
\end{figure}

\textbf{Analytic-2:}
Top companies that use ``Tableau" and ``MongoDB" technologies are shown in Fig. \ref{fig1:1} and top companies which hire people with ``master degree" and have ``Java" programming skills are shown in Fig. \ref{fig2:2}. 

\section{Conclusions and Future Work}
In this paper we have discussed a methodology that enhances the job searching experience of users by adding skill set and company attribute filters. Job seekers can pinpoint jobs by filtering on relevant skill sets, technologies, number of employees, micro industries and many other attributes. The filtered jobs are ranked using a relevance score derived from a weighted combination of skill sets and external factors of companies as described in \ref{QRR}. We have also developed advanced analytics on the data as shown in \ref{Analy}. 

We can additionally enhance the analytics with a temporal aspect by analyzing collected jobs information over time and identifying trends in the skill set requirements for different industries. We can also add more filters such as  additional social attributes that identify companies' social presence on the web. We can also add higher level attributes that are derived from a combination of existing attributes. For example, a higher level attribute such as ``Marketing Sophistication'' can be derived by a combination of social presence and advertising technologies attributes.


\section*{Acknowledgment}
We would like to thank Everstring for generously providing the datasets needed for our research and assisting us through out this project and being a source of motivation and knowledge. We would also like to thank Everstring's HPC (High Performance Computing) Group for providing us with constant technical support over their Hadoop cluster.

\end{document}